\documentstyle[11pt]{article}
\begin{document}

\title{Velocity Curve Analysis of the Spectroscopic Binary
   Stars V373 Cas, V2388 Oph, V401 Cyg, GM Dra, V523 Cas, AB And, and HD 141929 by
   Artificial Neural Networks}
\author{K. Karami$^{1,2}$,\thanks{E-mail: KKarami@uok.ac.ir}\\
K. Ghaderi${^1}$,\thanks{E-mail: K.Ghaderi.60@gmail.com}\\R. Mohebi${^1}$,\thanks{E-mail: rozitamohebi@yahoo.com}\\
R. Sadeghi${^3}$,\thanks{E-mail: rsadeghi@uok.ac.ir}\\
M.M. Soltanzadeh${^1}$,\thanks{E-mail:
msoltanzadeh@uok.ac.ir}\\$^{1}$\small{Department of Physics,
University of Kurdistan, Pasdaran St., Sanandaj,
Iran}\\$^{2}$\small{Research Institute for Astronomy
$\&$ Astrophysics of Maragha (RIAAM), Maragha, Iran}\\
$^{3}$\small{Department of Chemistry, University of Kurdistan,
Pasdaran St., Sanandaj, Iran}}

\maketitle

\begin{abstract}
We used an Artificial Neural Network (ANN) to derive the orbital
parameters of spectroscopic binary stars. Using measured radial
velocity data of seven double-lined spectroscopic binary systems
V373 Cas, V2388 Oph, V401 Cyg, GM Dra, V523 Cas, AB And, and HD
141929, we found corresponding orbital and spectroscopic elements.
Our numerical results are in good agreement with those obtained by
others using more traditional methods.
\end{abstract} \noindent{Key words.~~~stars: binaries: eclipsing --- stars: binaries: spectroscopic}
%-----------------------------------------------------------------------------------------------

\section{Introduction}
%Please see the PASA Style Guide for help with correct layout for your manuscript.
%Examples of tables and figures are given below.
\label{intro} Determining the orbital elements of binary stars
helps us to obtain fundamental information, such as the masses and
radii of individual stars, that has an important role in
understanding the present state and evolution of many interesting
stellar objects. Analyzing of both light and radial velocity
(hereafter RV) curves, derived from photometric and spectroscopic
observations, respectively, yields a complete set of basic
absolute parameters.

There are different methods to determine the orbit of a
spectroscopic binary from its RV curve. Lehmann-Filh\'{e}s (1894)
introduced a geometrical method to determine the orbital elements
from the geometrical properties of the RV curve, especially its
maxima and minima. The method of Lehmann-Filh\'{e}s has been found
to be very useful, and little, if any, longer than other methods,
providing a planimeter is used. This method also gives, for
improving the final solution of the orbit, the differential
corrections to the preliminary elements using the form of the
equations of condition obtained by the method of least squares
(see also Petrie 1960). Sterne (1941) described two forms of least
square solutions. The first form is suitable for all orbits except
those with very small eccentricities. The second form is
particularly suitable for orbits having very small eccentricities
(see also Petrie 1960). Karami $\&$ Teimoorinia (2007) introduced
a new non-linear least squares velocity curve analysis technique
for spectroscopic binary stars. Their method was applicable to
orbits of all eccentricities and inclination angles and the time
consumed was considerably less than the method of
Lehmann-Filh\'{e}s. They showed the validity of their new method
to a wide range of different types of binary. See Karami $\&$
Mohebi (2007a,b) and Karami et al. (2008).

In the present paper, we use an Artificial Neural Network (ANN) to
find the optimum match to the four parameters of the RV curves of
the seven double-lined spectroscopic binary systems: V373 Cas,
V2388 Oph, V401 Cyg, GM Dra, V523 Cas, AB And, and HD 141929. Our
aim is to show the validity of our new method for a wide range of
different binary types.

The spectral type of the primary and secondary component of V373
Cas is B0.5II and B4III, respectively. The mean effective
temperature is 22000$~\rm K$ and 18000$~\rm K$ for the primary and
secondary components. The angle of inclination is $\sim60^\circ$
with a period of 13.4 days (Hill $\&$ Fisher 1987). V2388 Oph is
very close visual binary. The spectral type is F3 V with a
relatively long period of 0.802 days. The system appears to be one
of the most luminous among currently known contact binaries.
Orbital inclination angle is $90^\circ$ (Rucinski et al. 2002).
V401 Cyg appears to be a rather typical contact system. The
orbital period is 0.582714 days (Rucinski et al. 2002). GM Dra is
a contact system of the W-type, with spectral type F5V and period
0.338741 days (Rucinski et al. 2002). V523 Cas is one of the
faintest known contact binaries. The spectral type is K4V and the
period is 0.233693 days (Rucinski et al. 2003). AB And is a
contact binary with spectral type G8V and period 0.3318919 days
(Pych et al. 2004). HD 141929 is a double-lined spectroscopic
binary with a period of 49.699 days. The orbit is eccentric
($e=0.393$). The effective temperature for both components is
estimated to $9500\pm250~\rm K$. Both components have the same
spectral type A0/1V, however, the secondary is rotating slower
than the primary and the inclination of the orbit is about
11$^\circ$ (Carrier 2002).

%\section{RV curve parameters estimation by the ANNs}
Following Smart (1990), the radial velocity of a star in a binary
system is defined as follows
\begin{eqnarray}
{\rm RV}=V_{\rm cm}+K[\cos(\theta+\omega)+e\cos\omega],\label{Vr}
\end{eqnarray}
where $V_{\rm cm}$ is the radial velocity of the center of mass of
system with respect to the sun. Where $\theta$, $\omega$ and $e$
are the angular polar coordinate (true anomaly), the longitude of
periastron and the eccentricity, respectively. Also
\begin{eqnarray}
K=\Large \frac{2\pi}{P}\frac{a\sin i}{\sqrt{1-e^2}},\label{K}
\end{eqnarray}
where $P$ is the period of motion, $a$ is the semimajor axis of
the orbit and inclination $i$ is the angle between the line of
sight and the normal of the orbital plane.

From Smart (1990) for a small eccentricity, $e\ll 1$, the true
anomaly, $\theta$, can be expressed in terms of the photometric
phase, $\phi$, as follows
\begin{eqnarray}
\theta=2\pi\phi&+&(2e-\frac{e^3}{4})\sin(2\pi\phi)+\frac{5}{4}e^2\sin(4\pi\phi)\nonumber \\
&+&\frac{13}{12}e^3\sin(6\pi\phi)+O(e^4).\label{expand}
\end{eqnarray}

\begin{figure}
\includegraphics{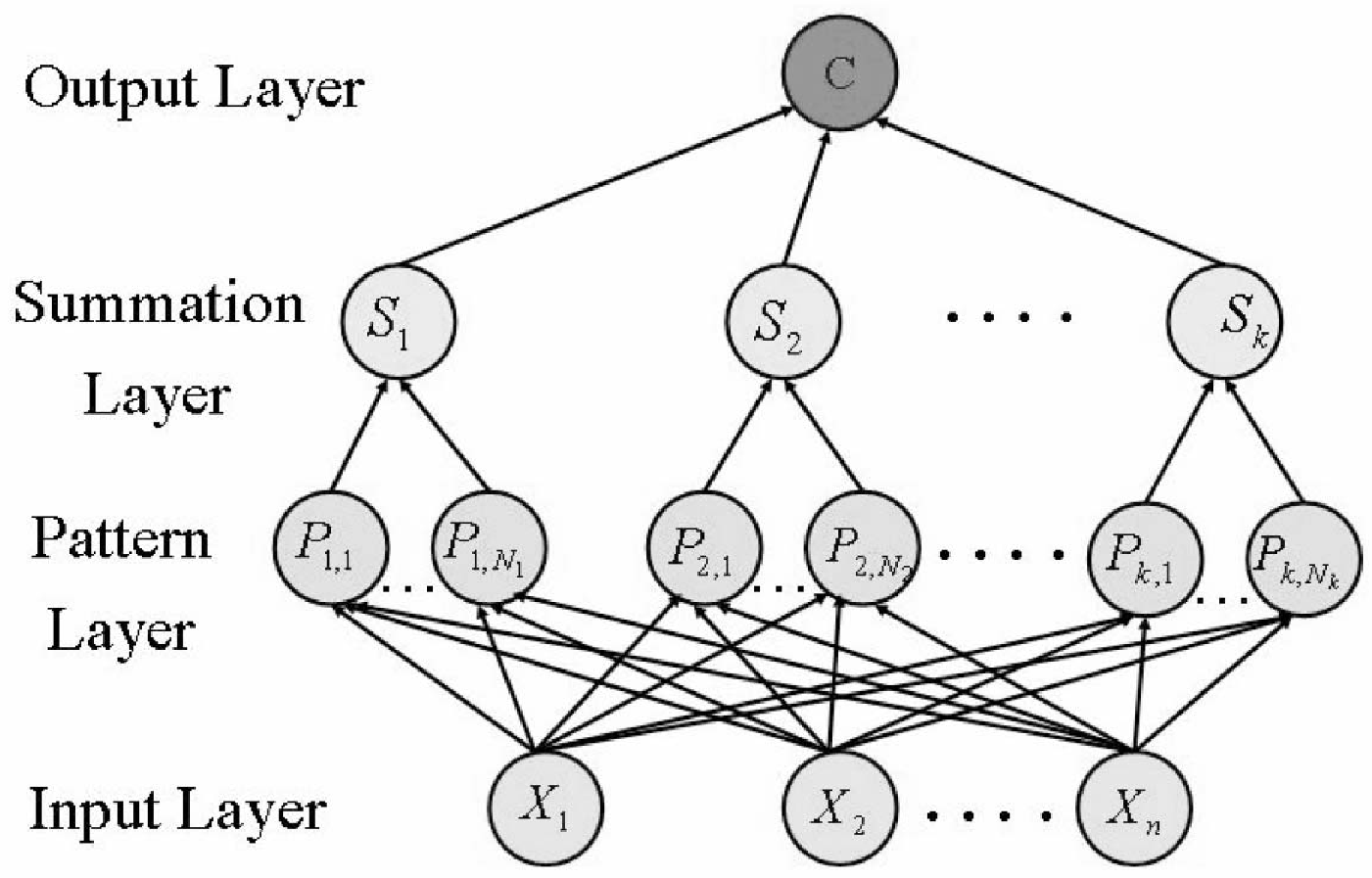}
      \vspace{6.3cm}
\caption[]{Schematic of a typical probabilistic neural network
given by Bazarghan et al. (2008).}
         \label{ANN}
   \end{figure}

Here we apply the ANN method to estimate the four orbital
parameters, $V_{\rm cm}$, $K$, $e$ and $\omega$ of the RV curve in
Eq. (\ref{Vr}). ANNs have become a popular tool in almost every
field of science. In recent years, ANNs have been widely used in
astronomy for applications such as star/galaxy discrimination,
morphological classification of galaxies, and spectral
classification of stars (see Bazarghan et al. 2008 and references
therein). Following Bazarghan et al. (2008), we employ
Probabilistic Neural Networks (PNNs). An example of a PNN is shown
in Fig. \ref{ANN}. This network has been investigated in ample
details by Bazarghan et al. (2008).

In this work, for the identification of the observational RV
curves, the input vector in Fig. \ref{ANN}, $X = (X_1, X_2, ...,
X_n)$, is the fitted RV curve of a star with $36$ data points $(n
= 36)$. This number is enough to cover the whole of RV curve. The
network is first trained to classify RV curves corresponding to
all the possible combinations of $V_{\rm cm}$ , $K$, $e$ and
$\omega$. For this we synthetically generate RV curves given by
Eq. (\ref{Vr}). We generate one RV curve for each combination of
the parameters:
\begin{itemize}
\item $-100\leq V_{\rm cm} \leq 100$ in steps of 1;
\item $1 \leq K \leq 300$ in steps of 1;
\item $0 \leq e \leq 1$ in steps of 0.001;
\item $0 \leq \omega \leq 360^\circ$ in steps of 5.
\end{itemize}

Note that from Petrie (1960), one can guess $V_{\rm cm}$, $K$ and
$e$ from a RV curve. This enable us to limit the range of
parameters around their initial guesses. This gives a set of
pattern groups, one group for each combination of $V_{\rm cm}$,
$K$, $e$ and $\omega$. Each pattern group, $k$, is characterized
by $N_k$ Gaussian functions (see Bazarghan et al. 2008). For each
system, we first fit a curve on the observational RV data. Then
using the fitted curve, the RV is computed in $36$ photometric
phases. When a observational RV curve of an unknown classification
is fed to the network, the summation layer of the network computes
the probability functions $S_k$ of each class. Finally at the
output layer we have C, the value with the highest probability
(see again Fig. \ref{ANN}).
%-----------------------------------------------------------------------------------------------
\section{Numerical Results}
\begin{figure}
\includegraphics{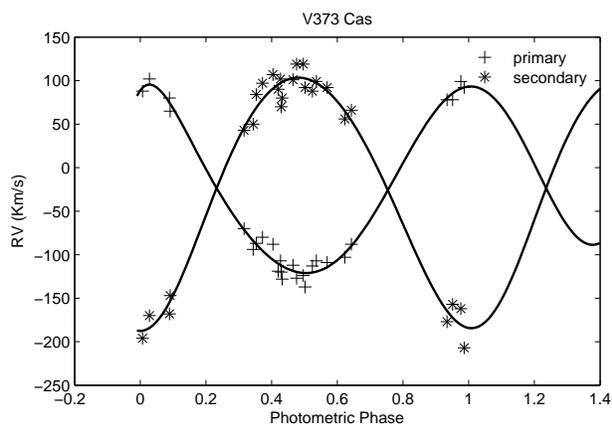}
      \vspace{8.6cm}
      \caption[]{Radial velocities of the primary and secondary component of V373 Cas
      plotted against the photometric phase. The observational data are from Hill $\&$ Fisher
      (1987).}
         \label{V373-RV}
   \end{figure}

\begin{figure}
\includegraphics{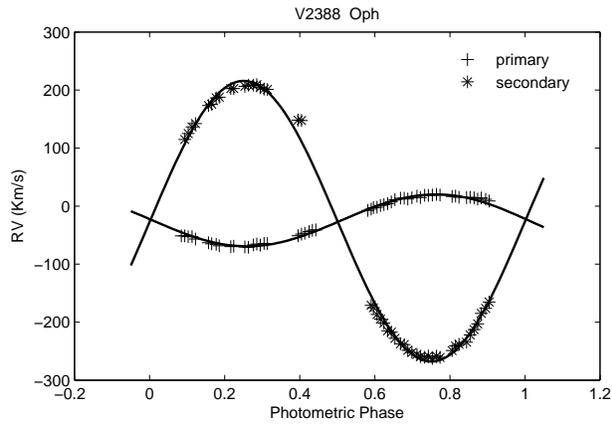}
      \vspace{8.7cm}
      \caption[]{Same as Fig. \ref{V373-RV}, for V2388 Oph. The observational data are from Rucinski et al.
       (2002).}
         \label{V2388-RV}
   \end{figure}

\begin{figure}
\includegraphics{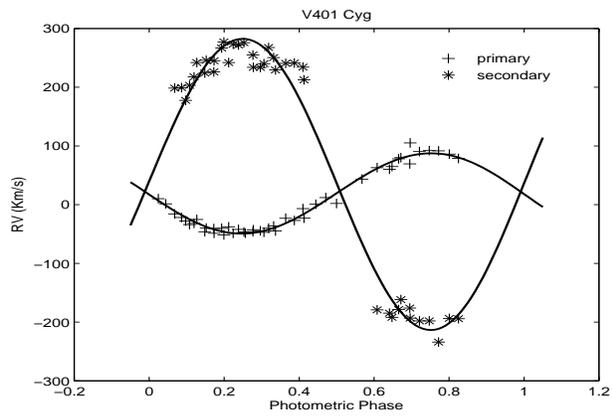}
      \vspace{10.5cm}
      \caption[]{Same as Fig. \ref{V373-RV}, for V401 Cyg. The observational data are from Rucinski et
      al. (2002).}
         \label{V401-RV}
   \end{figure}

\begin{figure}
\includegraphics{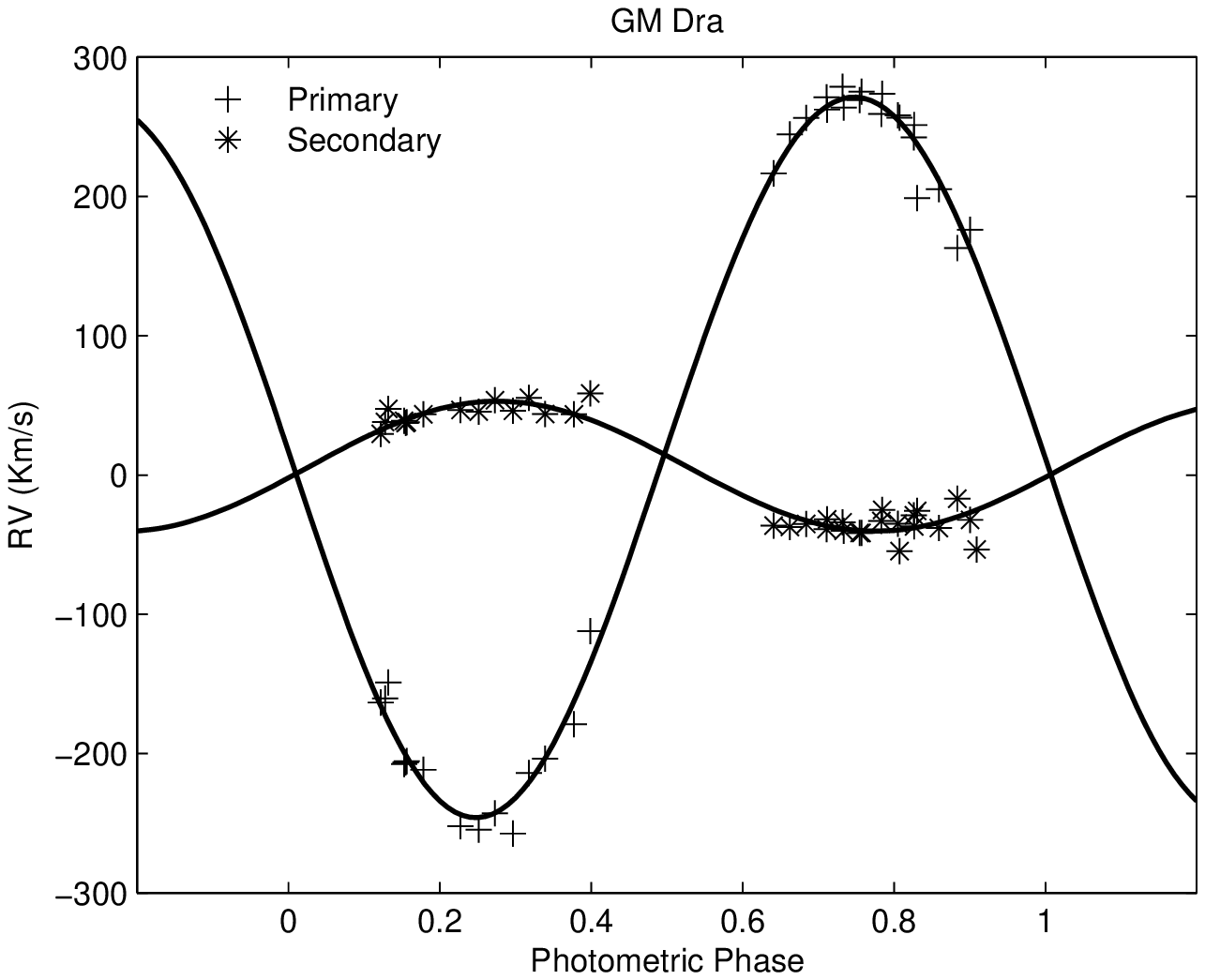}
      \vspace{9.5cm}
      \caption[]{Same as Fig. \ref{V373-RV}, for GM Dra.
      The observational data are from Rucinski et al. (2002).}
         \label{GM_Dra_RV}
   \end{figure}

\begin{figure}
\includegraphics{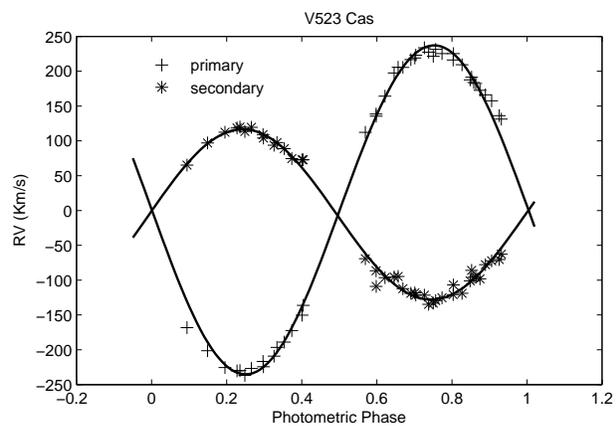}
      \vspace{8.5cm}
      \caption[]{Same as Fig. \ref{V373-RV}, for V523 Cas. The observational data are from Rucinski et al. (2003).
      }
         \label{V523-RV}
   \end{figure}

\begin{figure}
\includegraphics{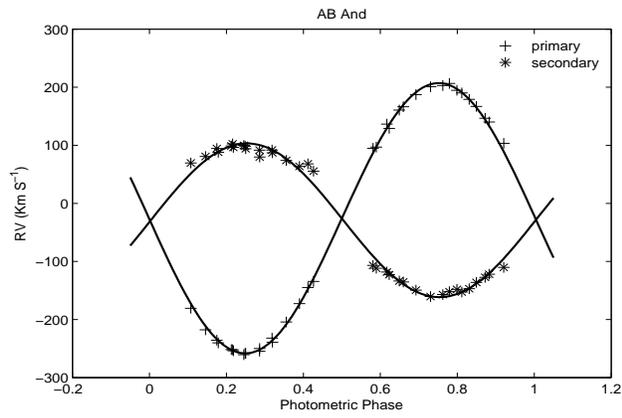}
      \vspace{10cm}
      \caption[]{Same as Fig. \ref{V373-RV}, for AB And. The observational data are from Pych et al.
      (2004).}
         \label{AB-RV}
   \end{figure}

\begin{figure}
\includegraphics{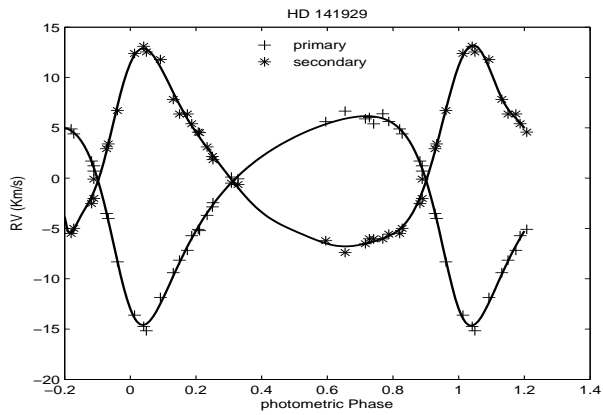}
      \vspace{9cm}
      \caption[]{Same as Fig. \ref{V373-RV}, for HD 141929.
      The observational data are from Carrier
      (2002).}
      \label{HD-141929-RV}
\end{figure}

Here PNN is used as a tool to derive the orbital parameters of the
seven different double-lined spectroscopic systems V373 Cas, V2388
Oph, V401 Cyg, GM Dra, V523 Cas, AB And, and HD 141929. Using
measured RV data of the two components of these systems obtained
by Hill \& Fisher (1987) for V373 Cas, Rucinski et al. (2002) for
V2388 Oph, V401 Cyg and GM Dra, Rucinski et al. (2003) for V523
Cas, Pych et al. (2004) for AB And, and Carrier (2002) for HD
141929, the fitted velocity curves are plotted in terms of the
photometric phase in Figs. \ref{V373-RV} to \ref{HD-141929-RV}.

The orbital parameters obtaining from the ANN for V373 Cas, V2388
Oph, V401 Cyg, GM Dra, V523 Cas, AB And, and HD 141929 are
tabulated in Tables \ref{V373-Orbit}, \ref{V2388-Orbit},
\ref{V401-Orbit}, \ref{GM-Dra-Orbit}, \ref{V523-Orbit},
\ref{AB-Orbit} and \ref{HD-141929-Orbit}, respectively. Tables
show that the results are in good accordance with the those
obtained by aforementioned authors. Tables \ref{V523-Orbit} and
\ref{AB-Orbit} show that the results of eccentricities for V523
Cas and AB And are not significantly different from zero. A Monte
Carlo analysis clears that they should really be assigned circular
orbits.

The combined spectroscopic elements including $m_p\sin^3i$,
$m_s\sin^3i$, $(a_p+a_s)\sin i$ and $m_p/m_s$ are calculated by
substituting the estimated parameters $K$, $e$ and $\omega$ into
Eqs. (3), (15) and (16) in Karami $\&$ Teimoorinia (2007). The
results obtained for the seven systems are tabulated in Tables
\ref{V373-Combined}, \ref{V2388-Combined}, \ref{V401-Combined},
\ref{GM-Dra-Combined}, \ref{V523-Combined}, \ref{AB-Combined} and
\ref{HD-141929-Combined} show that our results are in good
agreement with the those obtained by aforementioned authors.

Note that the errors in the tables of orbital parameters and
combined spectroscopic elements in Karami \& Mohebi (2009) seem to
be unrealistic errors. But they are indeed the standard errors
which are obtained from the nonlinear least squares of Eq. (14) in
Karami \& Teimoorinia (2007). The errors of observational RV data
are not included in them. Following Lucy $\&$ Sweeney (1971) the
meaningful errors $\sigma_i;~i=1,2,3,4$ corresponding to the
orbital elements $V_{\rm cm},K,e,\omega$, respectively, can be
obtained from the inverse of the error diagonal matrix $A$ as
$\sigma_i^2=\sigma^2A_{ii}^{-1}$ with $A_{11}=N$,
$A_{22}=A_{11}/2$, $A_{33}=NK^2/2$ and $A_{44}=(\pi
e/180)^2A_{33}$. Where $\sigma$ is the standard error of an
observation of average weight. Using Eqs. (3), (15) and (16) in
Karami $\&$ Teimoorinia (2007), the errors of combined
spectroscopic elements are obtained from the errors of orbital
parameters.
%-----------------------------------------------------------------------------------------------
\section{Conclusions}
An Artificial Neural Network to derive the orbital elements of
spectroscopic binary stars is applied. This method is applicable
to orbits of all eccentricities and inclination angles of
different types of binaries. In this method the time consumed is
considerably less than the method of Lehmann-Filh\'{e}s and even
less than the non-linear regression method introduced by Karami
$\&$ Teimoorinia (2007). It is possible to make adjustments in the
elements before the final result is obtained. There are some
cases, for which the geometrical methods are inapplicable, and in
these cases the present one may be found useful. One such case
would occur when observations are incomplete because certain
phases could not have been observed. Another case in which this
method is useful is that of a star attended by two dark companions
with commensurable periods. In this case the resultant velocity
curve may have several unequal maxima and the geometrical methods
fail altogether.

Using the measured RV data of V373 Cas, V2388 Oph, V401 Cyg, GM
Dra, V523 Cas, AB And, and HD 141929 given by aforementioned
authors, we find the orbital elements of these systems by the PNN.
Our numerical results shows that the results obtained for the
orbital and spectroscopic parameters are in good agreement with
those obtained by others using more traditional methods.
%-----------------------------------------------------------------------------------------------
\section*{Acknowledgments} The authors wish to thank Graham Hill who his elaborate
and meticulous comments has significantly improved the content and
the presentation of the paper. This work has been supported
financially by Research Institute for Astronomy $\&$ Astrophysics
of Maragha (RIAAM), Maragha, Iran.
%-----------------------------------------------------------------------------------------------

%-----------------------------------------------------------------------------------------------
\clearpage
%-----------------------------------------------------------------------------------------------
%\clearpage
%-----------------------------------------------------------------------------------------------
%\clearpage
%----------------------------------------------------------------------------------------------
%\clearpage
%-----------------------------------------------------------------------------------------------
%\clearpage
%-----------------------------------------------------------------------------------------------
%\clearpage
%----------------------------------------------------------------------------------------------
%\clearpage
%-----------------------------------------------------------------------------------------------
%\clearpage
%---------------------------------------------------------------------------------------------
\clearpage
\begin{table}
\centering\caption[]{Orbital parameters of V373 Cas.}
%\begin{center}
\begin{tabular}{lcccc}  \hline\noalign{\smallskip}
&This Paper  &Karami \& Mohebi (2009)& Hill $\&$ Fisher (1987)\\
\hline\noalign{\smallskip}
{\bf Primary} &   \\
$V_{\rm cm}\left(kms^{-1}\right)$ &$-25\pm1$& $-25.14\pm0.76$ & $-24.5\pm2$ \\
 $K_p\left(kms^{-1}\right)$& $110\pm1$& $109.52\pm0.22$ & $106.7\pm2.7$  \\
 $e$& $0.085\pm0.001$&$0.0972\pm0.0002$ & -- \\
$\omega(^\circ)$&$355\pm5$& $344.5\pm0.3$&--\\
{\bf Secondary} & \\
$V_{\rm cm}\left(kms^{-1}\right)$ & $-25\pm1$&$-25.14\pm0.76$ &--   \\
$K_s\left(kms^{-1}\right)$& $146\pm1$& $145.53\pm0.02$ &--   \\
$e$&$e_s=e_p$& $e_s=e_p$ &--\\
$\omega(^\circ)$&$175\pm5$&$w_s=w_p-180^\circ$&--\\\hline
\end{tabular}
\label{V373-Orbit}\\
\end{table}
%------------------------------------------------------------------------------------------------
\begin{table}
\centering\caption[]{Combined spectroscopic elements of V373 Cas.}
\begin{tabular}{lcccc}\hline\noalign{\smallskip}
 Parameter  & This Paper &Karami \& Mohebi (2009) & Hill $\&$ Fisher (1987) \\\hline\noalign{\smallskip}
$m_p\sin^3i/M_\odot$ &$13.1\pm0.3$& $12.98\pm0.03$&$12.6\pm0.2$\\
$m_s\sin^3i/M_\odot$ &$9.9\pm0.2$& $9.77\pm0.04$ &$9.3\pm0.2$ \\
$\left(a_p+a_s\right)\sin i/R_\odot$ &$67.5\pm0.5$& $67.3\pm0.1$ &$66.1\pm0.9$ \\
$m_p/m_s$ &$1.32\pm0.02$& $1.33\pm0.03$ & $1.35\pm0.04$ \\
\hline\noalign{\smallskip}
\end{tabular}\\
%\end{center}
\label{V373-Combined}
\end{table}
%------------------------------------------------------------------------------------------------
\begin{table}
\centering\caption[]{Same as Table \ref{V373-Orbit}, for V2388
Oph.}
%\begin{center}
\begin{tabular}{lcccc}  \hline\noalign{\smallskip}
&This Paper  &Karami \& Mohebi (2009)& Rucinski et al. (2002)\\
\hline\noalign{\smallskip}
{\bf Primary} &   \\
$V_{\rm cm}\left(kms^{-1}\right)$ &$-25\pm1$& $-25.35\pm0.88$ & $-25.88\pm0.52$ \\
 $K_p\left(kms^{-1}\right)$& $45\pm1$& $44.71\pm0.01$ & $44.62\pm0.48$  \\
 $e$&$0.008\pm0.001$& $0.006\pm0.001$ & -- \\
$\omega(^\circ)$&$285\pm5$& $276\pm2$&--\\
{\bf Secondary} & \\
$V_{\rm cm}\left(kms^{-1}\right)$ &$-25\pm1$& $-25.35\pm0.88$ &$-25.88\pm0.52$   \\
$K_s\left(kms^{-1}\right)$& $242\pm1$& $241.99\pm0.03$ &$240.22\pm0.98$   \\
$e$& $e_s=e_p$&$e_s=e_p$&--\\
$\omega(^\circ)$&$105\pm5$&$w_s=w_p-180^\circ$&--\\\hline
\end{tabular}
\label{V2388-Orbit}\\
\end{table}
%------------------------------------------------------------------------------------------------
\begin{table}
\centering\caption[]{Same as Table \ref{V373-Combined}, for V2388
Oph.}
\begin{tabular}{lcccc}\hline\noalign{\smallskip}
 Parameter  & This Paper &Karami \& Mohebi (2009) & Rucinski et al. (2002) \\\hline\noalign{\smallskip}
$m_p\sin^3i/M_\odot$ &$1.66\pm0.03$& $1.653\pm0.001$& --\\
$m_s\sin^3i/M_\odot$ &$0.31\pm0.01$& $0.306\pm0.001$ &-- \\
$\left(a_p+a_s\right)\sin i/R_\odot$ &$4.55\pm0.03$& $4.545\pm0.001$ &-- \\
$m_s/m_p$ &$0.187\pm0.005$& $0.185\pm0.001$ & $0.186\pm0.002$ \\
$\left(m_p+m_s\right)\sin^3i/M_\odot$&$1.97\pm0.04$&$1.959\pm0.002$&$1.926\pm0.030$\\
\hline\noalign{\smallskip}
\end{tabular}\\
%\end{center}
\label{V2388-Combined}
\end{table}
\label{lastpage}
%------------------------------------------------------------------------------------------------
\clearpage
\begin{table}
\centering\caption[]{Same as Table \ref{V373-Orbit}, for V401
Cyg.}
%\begin{center}
\begin{tabular}{lcccc}  \hline\noalign{\smallskip}
&This Paper  &Karami \& Mohebi (2009)& Rucinski et al. (2002)\\
\hline\noalign{\smallskip}
{\bf Primary} &   \\
$V_{\rm cm}\left(kms^{-1}\right)$ &$27\pm1$& $27.24\pm0.15$ & $25.53\pm2.14$ \\
 $K_p\left(kms^{-1}\right)$& $69\pm1$& $68.46\pm0.02$ & $72.23\pm2.43$  \\
 $e$&$0.006\pm0.001$& $0.0046\pm0.0001$ & -- \\
$\omega(^\circ)$&$185\pm5$& $181\pm5$&--\\
{\bf Secondary} & \\
$V_{\rm cm}\left(kms^{-1}\right)$ &$27\pm1$& $27.24\pm0.15$ & $25.53\pm2.14$   \\
$K_s\left(kms^{-1}\right)$& $248\pm1$& $247.99\pm0.02$ &$249.13\pm4.53$   \\
$e$& $e_s=e_p$&$e_s=e_p$ &--\\
$\omega(^\circ)$&$5\pm5$&$w_s=w_p-180^\circ$&--\\
\hline\noalign{\smallskip}
\end{tabular}\\
%\end{center}
\label{V401-Orbit}
\end{table}
%---------------------------------------------------------------------------------------------
\begin{table}
\centering\caption[]{Same as Table \ref{V373-Combined}, for V401
Cyg.}
\begin{tabular}{lccc}\hline\noalign{\smallskip}
 Parameter  & This Paper &Karami \& Mohebi (2009)& Rucinski et al. (2002)  \\\hline\noalign{\smallskip}
$m_p\sin^3i/M_\odot$ &$1.50\pm0.02$& $1.499\pm0.001$&--\\
$m_s\sin^3i/M_\odot$ &$0.42\pm0.01$& $0.414\pm0.001$ &-- \\
$\left(a_p+a_s\right)\sin i/R_\odot$ &$3.65\pm0.02$& $3.643\pm0.001$ &-- \\
$m_s/m_p$ &$0.278\pm0.005$& $0.276\pm0.001$ & $0.290\pm0.011$ \\
$\left(m_p+m_s\right)\sin^3i/M_\odot$&$1.92\pm0.03$&$1.913\pm0.002$&$2.008\pm0.130$\\
\hline\noalign{\smallskip}
\end{tabular}\\
%\end{center}
\label{V401-Combined}
\end{table}
%-----------------------------------------------------------------------------------------------
\begin{table}
\centering\caption[]{Same as Table \ref{V373-Orbit}, for GM Dra.}
%\begin{center}
\begin{tabular}{lccc}  \hline\noalign{\smallskip}
 &This Paper  &Karami \& Mohebi (2007a)& Rucinski et al. (2002) \\  \hline\noalign{\smallskip}
{\bf Primary} &   \\
$V_{\rm cm}\left(kms^{-1}\right)$ &$10\pm 1$& $10.55\pm0.59$ & $9.12\pm1.63$  \\
 $K_p\left(kms^{-1}\right)$&$258\pm1$&  $258.67\pm0.59$ & $258.72\pm2.66$  \\
 $e$&$0.005\pm0.001$& $0.008\pm0.004$ & --- \\
$\omega(^\circ)$&$165\pm5$& $175\pm2$ &---\\
{\bf Secondary} & \\
$V_{\rm cm}\left(kms^{-1}\right)$ &$10\pm 1$& $10.55\pm0.59$ & $9.12\pm1.63$   \\
 $K_s\left(kms^{-1}\right)$&$46\pm1$&  $46.59\pm0.98$ &$46.69\pm1.74$   \\
 $e$&$e_s=e_p$& $0.003\pm0.001$ &---\\
 $\omega(^\circ)$&$330\pm5$&$339\pm2$&--- \\  \hline\noalign{\smallskip}
\end{tabular}\\
%\end{center}
\label{GM-Dra-Orbit}
\end{table}
%------------------------------------------------------------------------------------------------
\begin{table} \centering\caption[]{Same as Table \ref{V373-Combined}, for GM Dra.}
%\begin{center}
\begin{tabular}{lccc}  \hline\noalign{\smallskip}
 Parameter  & This Paper&Karami \& Mohebi (2007a)& Rucinski et al. (2002)  \\  \hline\noalign{\smallskip}
 $m_p\sin^3i/M_\odot$ &$0.15\pm0.01$& $0.15\pm0.01$&--- \\
$m_s\sin^3i/M_\odot$ &$0.84\pm0.01$& $0.85\pm0.01$ &--- \\
$\left(a_p+a_s\right)\sin i/R_\odot$ &$2.03\pm0.01$& $2.04\pm0.01$ &--- \\
$m_p/m_s$ &$0.178\pm0.005$& $0.176\pm0.004$ & $0.180\pm0.002$ \\
$\left(m_p+m_s\right)\sin^3i/M_\odot$ &$0.99\pm0.02$& $1.00\pm0.02$&$1.002\pm0.042$\\
  \hline\noalign{\smallskip}
\end{tabular}\\
\label{GM-Dra-Combined}
\end{table}
%-----------------------------------------------------------------------------------------------
\clearpage
\begin{table}
\centering\caption[]{Same as Table \ref{V373-Orbit}, for V523
Cas.}
%\begin{center}
\begin{tabular}{lcccc}\hline\noalign{\smallskip}
&This Paper  &Karami \& Mohebi (2009)& Rucinski et al. (2003)\\
\hline\noalign{\smallskip}
{\bf Primary} &   \\
$V_{\rm cm}\left(kms^{-1}\right)$ &$-2\pm1$& $-2.31\pm0.71$ & $-2.54\pm0.90$ \\
 $K_p\left(kms^{-1}\right)$& $237\pm1$& $236.22\pm0.04$ & $235.95\pm1.41$  \\
 $e$&$0.002\pm0.001$& $0.0012\pm0.0002$ & -- \\
$\omega(^\circ)$&$190\pm5$& $190\pm11$&--\\
{\bf Secondary} & \\
$V_{\rm cm}\left(kms^{-1}\right)$ &$-2\pm1$& $-2.31\pm0.71$ & $-2.54\pm0.90$   \\
$K_s\left(kms^{-1}\right)$& $123\pm1$& $122.38\pm0.02$ &$121.64\pm1.14$   \\
$e$& $e_s=e_p$&$e_s=e_p$ &--\\
$\omega(^\circ)$&$10\pm5$&$w_s=w_p-180^\circ$&--\\\hline
\end{tabular}
\label{V523-Orbit}\\
\end{table}
%------------------------------------------------------------------------------------------------
\begin{table}
\centering\caption[]{Same as Table \ref{V373-Combined}, for V523
Cas.}
\begin{tabular}{lccc}\hline\noalign{\smallskip}
 Parameter  & This Paper &Karami \& Mohebi (2009) & Rucinski et al. (2003) \\\hline\noalign{\smallskip}
$m_p\sin^3i/M_\odot$ &$0.39\pm0.01$& $0.381\pm0.002$&--\\
$m_s\sin^3i/M_\odot$ &$0.74\pm0.01$& $0.736\pm0.001$ &-- \\
$\left(a_p+a_s\right)\sin i/R_\odot$ &$1.66\pm0.01$& $1.6557\pm0.0003$ &-- \\
$m_p/m_s$ &$0.52\pm0.01$& $0.5177\pm0.0002$ & $0.516\pm0.007$ \\
$\left(m_p+m_s\right)\sin^3i/M_\odot$&$1.13\pm0.02$&$1.117\pm0.003$&$1.11\pm0.24$\\\hline
\end{tabular}
\label{V523-Combined}\\
\end{table}
%-----------------------------------------------------------------------------------------------
\begin{table}
\centering \caption[]{Same as Table \ref{V373-Orbit}, for AB And.}
\begin{tabular}{lcccc}\hline\noalign{\smallskip}
 &This Paper  &Karami et al. (2008)& Pych et al. (2004)&\\
\hline\noalign{\smallskip}
 {\bf Primary} &   \\
$V_{\rm cm} \left(kms^{-1}\right)$ & $-27\pm 1$&$-27.26\pm 0.66$ & $-27.53\pm0.67$&  \\
 $K_p\left(kms^{-1}\right)$& $233\pm1$& $232.69\pm0.02$ & $232.88\pm0.83$ &  \\
 $e$& $0.002\pm0.001$&$0.00109\pm 0.00005$ &---&  \\
$\omega(^\circ)$&$220\pm5$& $231\pm 3$ &---& \\
{\bf Secondary} & \\
$V_{\rm cm} \left(kms^{-1}\right)$ &$-27\pm 1$& $-27.26\pm0.66$ & $-27.53\pm0.67$ &\\
 $K_s\left(kms^{-1}\right)$&$133\pm1$&  $132.43\pm0.01$ &$130.32\pm1.17$ &\\
 $e$&$e_s=e_p$& $e_s=e_p$ &---& \\
 $\omega(^\circ)$&$40\pm5$&$\omega_s=\omega_p-180^\circ$&---&
\\\hline\noalign{\smallskip}
\end{tabular}\\
\label{AB-Orbit}
\end{table}
%-----------------------------------------------------------------------------------------------
\begin{table}
\centering\caption[]{Same as Table \ref{V373-Combined}, for AB
And.}
\begin{tabular}{lccc}\hline\noalign{\smallskip}
 Parameter  & This Paper &Karami et al. (2008)& Pych et al. (2004) \\\hline\noalign{\smallskip}
 $m_p\sin^3i/M_\odot$ &$0.61\pm0.01$& $0.6071\pm 0.0001$ & ---\\
$m_s\sin^3i/M_\odot$ & $1.07\pm0.02$&$1.0668\pm 0.0002$ & ---\\
$\left(a_p+a_s\right)\sin i/R_\odot$ &$2.40\pm0.01$& $2.3943\pm 0.0002$ & ---\\
$m_p/m_s$ &$0.57\pm0.01$& $0.5691\pm 0.0001$ & $0.560\pm0.007$\\
$\left(m_p+m_s\right)\sin^3i/M_\odot$ &$
1.68\pm0.03$&$1.6739\pm0.0003$&$1.648\pm0.020$\\\hline\noalign{\smallskip}
\end{tabular}\\
\label{AB-Combined}
\end{table}
%-----------------------------------------------------------------------------------------------
\clearpage
\begin{table}
\centering
 \caption[]{Same as Table \ref{V373-Orbit}, for HD
141929.}
%\begin{center}
\begin{tabular}{lccc} \hline\noalign{\smallskip}
 &This Paper  &Karami \& Mohebi (2007b)& Carrier (2002) \\\hline\noalign{\smallskip}
{\bf Primary} &   \\
$V_{\rm cm}\left(kms^{-1}\right)$ &$-1\pm1$& $-0.44\pm0.12$ & $-0.33\pm0.08$  \\
 $K_p\left(kms^{-1}\right)$&$11\pm1$&  $10.38\pm0.01$ & $10.58\pm0.16$  \\
 $e$& $0.398\pm0.001$&$0.391\pm0.001$ & $0.393\pm0.008$ \\
$\omega(^\circ)$&$140\pm5$& $148.04\pm0.34$ &$145.7\pm1.7$\\
{\bf Secondary} & \\
$V_{\rm cm}\left(kms^{-1}\right)$ &$-1\pm1$& $-0.44\pm0.12$ & $-0.33\pm0.08$   \\
 $K_s\left(kms^{-1}\right)$& $10\pm1$& $9.87\pm0.01$ &$9.95\pm0.17$   \\
 $e$&$e_s=e_p$& $0.389\pm0.001$ &$0.393\pm0.008$\\
 $\omega(^\circ)$&$320\pm5$&$324.03\pm0.34$&$325.7\pm1.7$ \\\hline\noalign{\smallskip}
\end{tabular}\\
%\end{center}
\label{HD-141929-Orbit}
\end{table}
%------------------------------------------------------------------------------------------------
\begin{table} \caption[]{Same as Table \ref{V373-Combined}, for HD
141929.} \centering
%\begin{center}
\begin{tabular}{lccc} \hline\noalign{\smallskip}
 Parameter  & This paper&Karami \& Mohebi (2007b) & Carrier (2002)  \\\hline\noalign{\smallskip}
 $m_p\sin^3i/M_\odot$ &$0.017\pm0.005$& $0.0163\pm0.0001$&$0.01681\pm0.00064$ \\
$m_s\sin^3i/M_\odot$ &$0.019\pm0.005$& $0.0171\pm0.0001$ &$0.01789\pm0.00067$ \\
$ a_p\sin i/10^6Km$ &$6.9\pm0.6$& $6.53\pm0.01$ &$6.65\pm0.11$ \\
$ a_s\sin i/10^6Km$ &$6.3\pm0.6$& $6.21\pm0.02$ &$6.25\pm0.11$ \\
$m_p/m_s$ &$0.9\pm0.2$& $0.953\pm0.004$ & $0.94\pm0.02$ \\
 \hline\noalign{\smallskip}
\end{tabular}\\
%\end{center}
\label{HD-141929-Combined}
\end{table}
%------------------------------------------------------------------------------------------------

\end{document}